\documentclass[%
 aip,
 amsmath,amssymb,
 reprint,%
]{revtex4-1}

\usepackage{graphicx}
\usepackage{dcolumn}
\usepackage{bm}

\usepackage[utf8]{inputenc}
\usepackage[T1]{fontenc}
\usepackage{mathptmx}
\usepackage[separate-uncertainty=true]{siunitx}
\usepackage{braket}
\usepackage[colorlinks=true,linkcolor=blue]{hyperref}

\begin{document}

\preprint{AIP/123-QED}

\title{Electromagnetic induction imaging with a scanning radio-frequency atomic magnetometer}
\author{Cameron Deans}
\altaffiliation[Current address:]{RAL Space, STFC - Rutherford Appleton Laboratory, Fermi Avenue, Harwell, Didcot, OX11 0QX, United Kingdom}
\author{Yuval Cohen}
\author{Han Yao}
\author{Benjamin Maddox}
\author{Antonio Vigilante}
\author{Ferruccio Renzoni}
\email{f.renzoni@ucl.ac.uk}
\affiliation{Department of Physics and Astronomy, University College London, Gower Street, London WC1E 6BT, United Kingdom}

\date{\today}

\begin{abstract}
We demonstrate electromagnetic induction imaging with an unshielded, portable radio-frequency atomic magnetometer scanning over the target object. This configuration satisfies standard requirements in typical applications, from security screening to medical imaging. The ability to scan the magnetometer over the object relies on  the miniaturization of the sensor head and on the active compensation of the ambient magnetic field. Additionally, a procedure is implemented to extract high-quality images from the recorded spatial dependent magnetic resonance. The procedure is shown to be effective in suppressing the detrimental effects of the spatial variation of the magnetic environment. 
\end{abstract}

\maketitle
Electromagnetic induction imaging (EMI) \cite{griffiths2001} is a non-contact imaging technique which allows the spatial mapping of the electromagnetic properties of an object. Applications in security \cite{brendan2015a} and medical imaging \cite{alzebaik1993,grazfeasibility,zol2010,wang2017} would greatly benefit from the inherently safe nature of the technique, due to the lack of ionising radiation, while its technical simplicity makes it easily deployable, thus ideal for rapid assessment of brain injuries. 
EMI relies on the induction of eddy currents in the target by a low-frequency magnetic field, and the measurements of the resulting secondary magnetic field. Conventional set-ups are based on pick-up coils for the readout of the secondary field. However, this approach is limited by the poor sensitivity of coils at low frequency, which has hindered the development of applications of EMI. The combination of EMI with ultra-sensitive atomic magnetometers (AMs) \cite{arne2013,arne2014} has unlocked the potential of the technique,  opening up a wealth of applications, from medical imaging \cite{luca2016,spie2016}, to security and surveillance \cite{marmugi2015spie,cameron2017,cameron2018b}, and industrial monitoring.

All experimental demonstrations to date \cite{cameron2016,arne2016,luca2019,jensen2019,luca2020,patrick2018,npl2019a,npl2019b} of electromagnetic induction imaging with atomic magnetometers (EMI-AM) rely on the displacement of the target object with respect to the fixed AM. Such arrangements are at odds with the requirements of imaging for many applications, where the ability to scan the sensor over the object is often needed. This is due to the technical difficulties associated with scanning AMs. First, recent realisations of EMI-AM rely on radio-frequency atomic magnetometers (RF-AM) \cite{savukov2005,budker} which typically require two laser beams at different frequency and a radio-frequency source, hence are more difficult to miniaturise
than, for example, coherent population trapping \cite{cpt} or free-induction decay \cite{fid} magnetometers. Second, operation of an atomic magnetometer at extreme sensitivity requires a controlled environment. 
Typically systems employ several layers of $\mu$-metal shielding to suppress magnetic noise. This limits the scope of practical applications to targets within the shields. In contrast, RF-AMs have been shown to retain their extreme sensitivity in unshielded environments. This is usually achieved by compensating stray magnetic fields at the position of the AM. However, this results in an RF-AM that is optimized to a fixed position. Hence, until now, the EMI-AM procedure is simplified by keeping the AM fixed while moving the target.

In this Letter we demonstrate EMI with an unshielded, portable RF-AM scanning over a fixed target object, thus satisfying the requirements for real-world applications.  Our  reported demonstration relies on two innovations. First, a compact RF-AM is realised, with the sensor head including the required lasers sources, the RF source as well as the magnetic field coils for the active compensation of stray magnetic field. Second, we note that the active compensation system used for stray magnetic fields cannot exactly cancel the magnetic environment. Thus a procedure is implemented  to reduce the detrimental effects of the residual spatial variations in the magnetic environment experienced by the sensor head while scanning over the object. 

\begin{figure}[ht]
	\includegraphics[width=\linewidth]{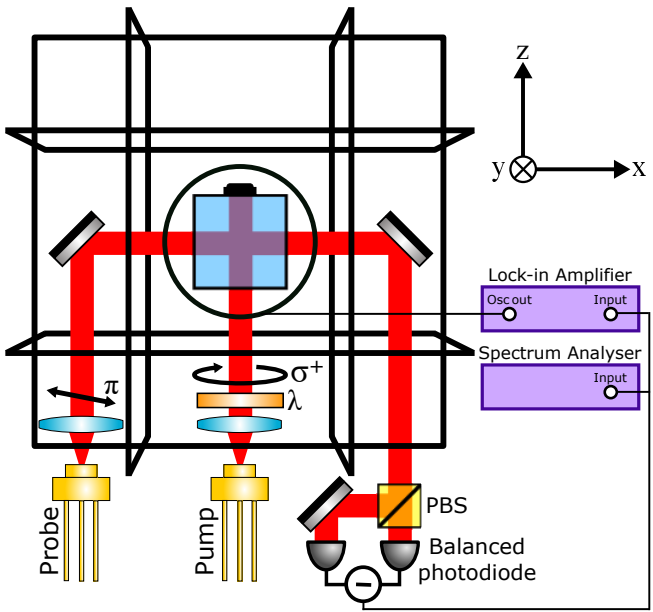}
	\centering
	\caption[]{Schematic overview of the magnetometer.} 
	\label{fig:setup_2}
\end{figure}

Our sensor head follows the  standard arrangement of an RF-AM, as sketched in Figure~\ref{fig:setup_2}. An atomic vapour is spin-polarized by optical pumping via a $\sigma^{+}$ laser beam in the presence of an applied bias magnetic field collinear to the laser beam. A perpendicular AC magnetic field ($\mathbf{B_{RF}}$) excites spin-coherences and produces a transverse atomic polarization. A linearly polarized probe laser beam is transmitted through the vapour, perpendicular to the pump beam. The atomic Larmor precession is mapped on the rotation of the probe plane of polarization which is measured by a polarimeter. The output is then interrogated by a lock-in amplifier (LIA) and a spectrum analyzer.

\begin{figure}[ht]
	\includegraphics[width=\linewidth]{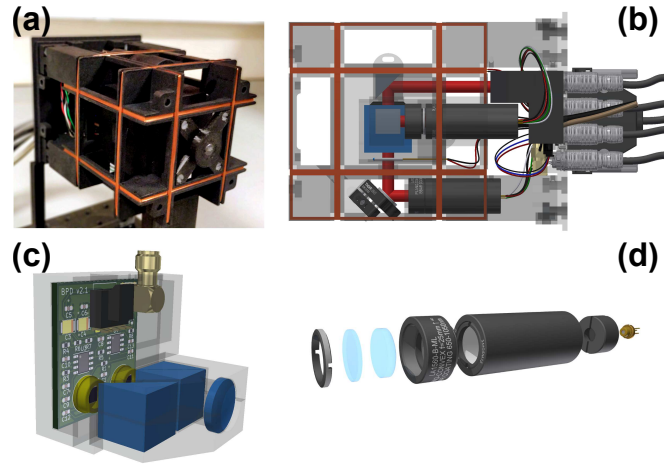}
	\centering
	\caption{(a) Photo of the sensor head, with the cover removed. (b) CAD sketch of the magnetometer showing the two lasers, beam path, coils, cell and electronic wiring. (c) The polarimeter. The probe enters after passing through the cell, where an F=\SI{20}{\milli\meter} lens focuses the beam for detection. A polarising beam splitter and a mirror direct the two separated components of polarisation towards the two Thorlabs FDS100 photodiodes, separated by 14 mm and mounted on a PCB. The same PBC includes the differential amplification circuit. An SMA jack allows for readout of the output signal. The total volume of the polarimeter is \SI{29.7}{\centi\meter\cubed}. (d) An exploded view of the Vixar I0-0795S-0000-BC06 VCSEL laser packaging. The chip output is collimated by an F=\SI{25}{\milli\meter} lens. For the pump a $\lambda/4$ waveplate is included to set the $\sigma^+$ polarization.
	 The whole package is \SI{18}{\milli\meter} in diameter and \SI{45}{\milli\meter} in length.}
	\label{fig:setup_1}
\end{figure}

The implementation of the sensor head is shown in Figure \ref{fig:setup_1}. A support in nylon, 3D printed via selective laser sintering (SLS),  holds all the optical and electronic parts in place, with a cover printed in the same material enclosing the sensor. A cubic glass cell of 25 mm side contains Rb vapour and 20 Torr of N$_2$ which acts both as a buffer gas and as a quenching gas. Initial tests were carried out with a gently heated isotopically enriched $^{87}$Rb vapour. However, a room temperature naturally occurring isotopic mixture of Rb was found to be sufficient for high-quality imaging, and was installed and used for subsequent measurements. Two Vixar  I0-0795S-0000-BC06  VCSEL lasers, with internal  thermoelectric coolers (TECs) and thermistors, are integrated in the sensor head and provide pumping and cooling light. Currents and temperatures are controlled by identical Thorlabs LDC200CV current supplies and Newport 325 temperature controller units. The lasers are tuned by varying the current and temperature. The pump laser is tuned in resonance with the 
$^{87}$Rb $5^2S_{1/2}, F = 1 \rightarrow 5^2P_{1/2}, F' = 2$ transition. 
The probe laser is first tuned in resonance with the  D$_1$ line $F=1 \to F^{\prime}$  manifold, and then finely tuned by optimising the atomic magnetic resonance used for sensing.  
These detunings are obtained with a current/temperature settings of 1.7 mA and 69.7$^\circ$ C for the probe and 2.0 mA and 70.4$^\circ$ C for the pump.

The sensor head also includes a system for the active compensation of spurious magnetic fields. A 3-axis fluxgate (Bartington MAG619) is placed near the cell and serves as an input for a feedback loop.
Three proportional-integral-derivate (PID) modules (Standford Research Systems SIM960) drive current in three Helmholtz coil pairs (35 turns, 0.2 mm diameter copper wire) surrounding the sensor.
The fluxgate has a bandwidth of DC - \SI{3}{\kilo\hertz}, limiting the bandwidth of the feedback loop, so ambient low-frequency magnetic noise is compensated for without affecting the applied radio-frequency driving the magnetometer. The same system of coils used for the compensation of the ambient magnetic field is used to generate the bias magnetic field required for the operation of the RF-AM. The system is designed to produce a maximum magnetic field of 1.4 G at 250 mA along each orthogonal axis within the cell volume. This value is large enough to cancel any ambient magnetic fields and provide a bias field in the desired operating range - up to 100 kHz - regardless of the orientation of the sensor head.

The sensor head also includes a miniaturized balanced polarimeter. It contains a focusing lens (as in the figure description), a polarizing beam splitter (PBS), a mirror and two photodiodes integrated on a printed circuit board (PCB) with the circuit for differential amplification.  The total volume of the polarimeter is \SI{29.7}{\centi\meter\cubed}. The dimensions and weight of the fully assembled head are $110 \times 110\times 145$ mm$^3$ and 1.49 kg.

\begin{figure}[htbp]
	\includegraphics[width=\linewidth]{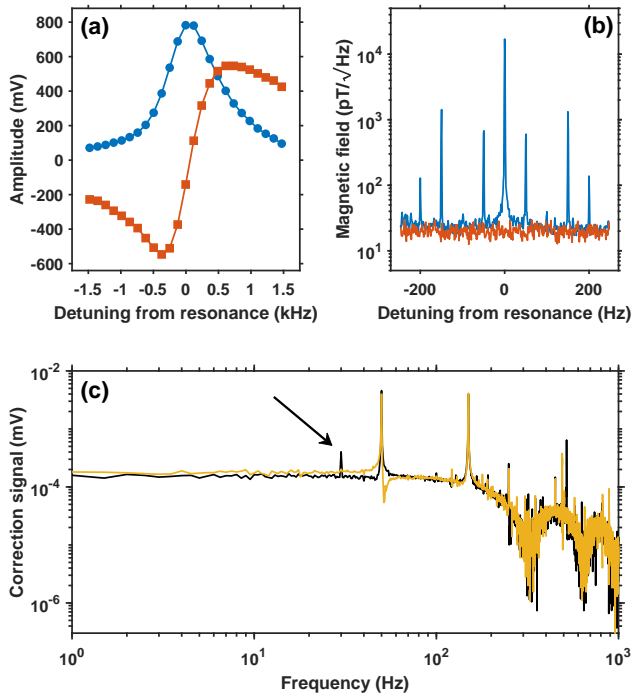}
	\centering
	\caption{(a) A typical atomic magnetic resonance from our system, with the in-phase (blue circles) and out-of phase (red squares) signals obtained via demodulation of the polarimeter signal with a lock-in amplifier. (b) Power spectrum of magnetometer signal,
	giving a signal to noise ratio of 915. (c) PID correction signal of the bias field stabilisation in two different configurations: one operating in the ambient magnetic field (yellow) and the other (black) in the presence of an applied \SI{30}{\hertz} oscillating magnetic field, indicated by an arrow. Data are for a gently heated ($\simeq 40^\circ$ C) isotopically enriched $^{87}$Rb vapour and 20 Torr of N$_2$.  }
	\label{fig:resonance}
\end{figure}

The performance of the atomic magnetometer is evaluated by determining its sensitivity.  For this, a \SI{17}{\nano\tesla} calibration field is applied. The RF field is scanned around resonance, and the polarimeter output is  demodulated by a lock-in amplifier. The measured atomic magnetic resonance is displayed in  Figure \ref{fig:resonance} (a), with the in-phase and out of phase lock-in amplifier outputs reported as a function of the detuning from resonance. The half-width at half-maximum (HWHM) of the atomic magnetic resonance is \SI{460}{\hertz}. 
Figure \ref{fig:resonance} (b) shows the power spectrum of the polarimeter signal when the same field used in Figure \ref{fig:resonance}(a) is applied, with the reported baseline noise level obtained with the RF switched off. A signal to noise ratio of 915 is extracted. From the presented data we derive an AC sensitivity of \SI{19}{\pico\tesla\per\sqrt\hertz} and a DC sensitivity of \SI{22}{\pico\tesla\per\sqrt\hertz}.

The active magnetic field compensation acts along the three directions, as any uncompensated ambient field would shift the magnetometer out of resonance. The most critical compensation is along the bias field.
We verified the effective operation of the feedback loop by applying a  \SI{30}{\hertz} oscillating magnetic field and observing the PID correction signal. The appearance of an additional resonance, as shown in Figure \ref{fig:resonance}(c), shows that the PID reacts to the \SI{30}{\hertz} field and acts to cancel it.


Imaging is performed by scanning the magnetometer over a target object held at a fixed position. The sensor head is mounted on a computer-controlled motorized XY stage. Target objects can be placed on a plastic sheet above the scanning magnetometer.

\begin{figure}[htbp]
	\includegraphics[width=\linewidth]{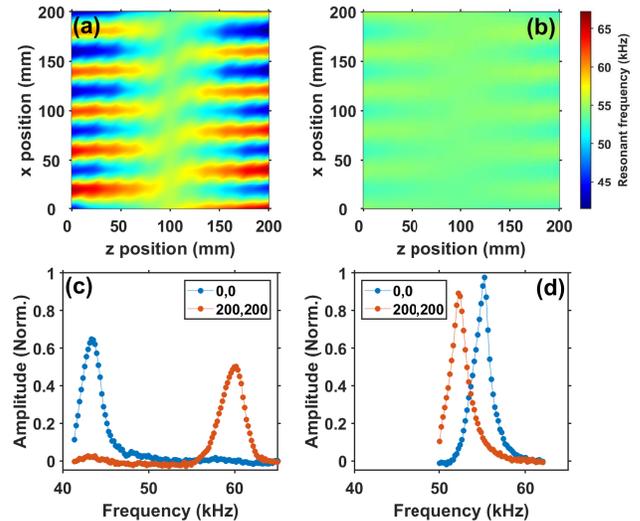}
	\centering
	\caption{Spatial map of the magnetic background. (a,b) Spatial dependence of the resonance frequency of the AM without (a) [with (b)] active compensation of the stray magnetic fields. 
	(c,d) Atomic magnetic resonance at the two opposite edges of the imaged region, without (c) [with (d)] active compensation of the ambient magnetic field. The amplitude of the resonance is normalised to the maximum amplitude recorded in the image with active feedback. Data are for naturally occurring $^{85,87}$Rb vapour and 20 Torr of N$_2$ at room temperature.}
	\label{fig:background_locked_unlocked}
\end{figure}

A first investigation was devoted to the study of the magnetic environment and specifically to its spatial dependence. An 11 $\times$ 11 pixel scan is taken in a 200 $\times$ 200 mm plane without any target object. At each position, the RF was scanned and the resonance frequency determined. Two different sets of measurements were taken, the first one without active stabilisation, which was then activated for the subsequent measurements. Results of this background mapping are presented in Figure \ref{fig:background_locked_unlocked}. Figure \ref{fig:background_locked_unlocked}(a) illustrates the background with the feedback loop left open. A \SI{25.6}{\kilo\hertz} variation in resonance frequency is seen across the image. For this measurement the output of the PID was manually adjusted to the value corresponding to the PID output at the centre of the image with feedback closed.
Figure \ref{fig:background_locked_unlocked}(b) shows the effect of active compensation, as obtained by closing the three feedback loops: the variation in resonance frequency  is reduced to \SI{3.2}{\kilo\hertz}. The detrimental effect of the spatial variation of the magnetic background is also highlighted in Figure \ref{fig:background_locked_unlocked}(c,d), where the magnetic resonances at opposite edges (position 0,0 mm and 200,200 mm) of the images are measured without/with active compensation in operation. Active stabilization not only reduces the variation in resonance frequency across the image, but also increases the resonance amplitude
 \cite{cameron2018c}. A consistent background is thus crucial for imaging of conductive objects.   While active compensation is essential, we will show in the following that additional procedures are required to obtain high-quality imaging with a scanning atomic magnetometer.

\begin{figure}[htbp]
	\includegraphics[width=\linewidth]{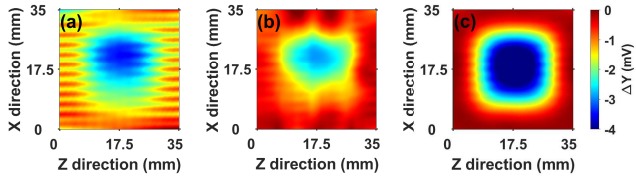}
	\centering
	\caption{Performance of resonance tracking and detuning for imaging. All images show Y and the colormaps are normalised to the scale on the right. All images are produced by using the same set of raw data generated by the lock-in amplifier. (a) The value of Y at 54kHz is plotted. The resonant frequency for the image varies from a minimum of 53.8 kHz to a maximum of 54.2 kHz, so the selected frequency corresponds to the central value. (b) Results obtained with resonance tracking. (c) Results obtained by introducing a detuning of +800 Hz from the tracked resonance.  Data are for naturally occurring $^{85,87}$Rb vapour and 20 Torr of N$_2$ at room temperature.
}
	\label{fig:resonance_tracking}
\end{figure}

EMI was performed by scanning the magnetometer over the target sample. For each position, the rf frequency is scanned around resonance, with the four outputs of the lock-in amplifier,
the in-phase response (X), the quadrature response (Y), the amplitude of the response (R), and its phase lag ($\Phi$), collected during the scan. The active compensation of the magnetic field is not sufficient to reduce the level of the spatial variation of the magnetic environment. This is visible in Figure \ref{fig:resonance_tracking}(a), where an image of a copper square is produced by plotting the value of $Y$ at the same frequency for all pixels.  Lines across the image are due to the resonance frequency changing at different positions, as seen in the background scan shown in Figure \ref{fig:background_locked_unlocked}. High-quality imaging was obtained by adopting the following procedure, consisting of two elements. First, the resonance was tracked in the imaging: for each position the magnetic resonance was fitted, and the total signal height (Y) at resonance was used, as opposed to at a set frequency. The improvement in imaging performance is visible in \ref{fig:resonance_tracking}(b). Second, the image is produced by introducing a detuning from resonance, and specifically taking the measurement for Y at +\SI{800}{\hertz} detuning from resonance, a procedure initially established for low-conductivity imaging in the configuration with a fixed RF-AM \cite{luca2019,luca2020}  This approach flattens the background further, producing a clear image of a copper square, as shown in Figure \ref{fig:resonance_tracking}(c) . 

\begin{figure}[htbp]
	\includegraphics[width=\linewidth]{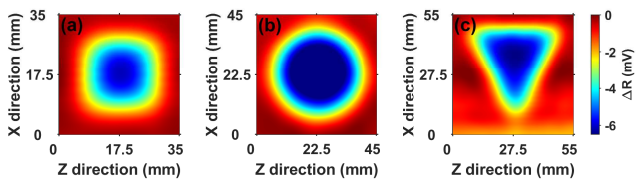}
	\centering
	\caption{Images of conductive objects. All image colormaps are normalised to the scale on the right. Part (a) and (b) are taken with 1.5mm step size, and part (c) is taken with 1.2mm step size.
	 Nearest-neighbour filtering helps with noise reduction in the image, with the data shown using a 3 nearest neighbour Gaussian filter. Data are for naturally occurring $^{85,87}$Rb vapour and 20 Torr of N$_2$ at room temperature.  }
	\label{fig:imaging}
\end{figure}

The imaging performance is illustrated in Figure \ref{fig:imaging}, which reports EMI images of a $25\times 25\times 1$ mm Copper square, a \SI{37}{\milli\meter} diameter \SI{2}{\milli\meter} thick aluminium circle,  and a \SI{50}{\milli\meter} base and height, \SI{2}{\milli\meter} thick aluminium isosceles triangle. All these images were taken with a scanning sensor head, active compensation of the magnetic field and the procedure including the tracking of the resonance and near-resonant imaging. 
Data shown are for R, but similar images are also produced for X, Y and $\Phi$. All objects are clearly imaged, with their shapes well resolved.  These images demonstrate the capability of performing EMI with a scanning radio-frequency magnetometer. We note that  none of the images presented in this work relies on background subtraction, further validating the use of the magnetometer in the field with no calibration.

In conclusion, in this work we demonstrated electromagnetic induction imaging with a radio-frequency atomic magnetometer scanning over  target objects. The demonstration relies on a portable sensor head, with integrated laser sources, RF source and active compensation of ambient magnetic fields, and a dedicated imaging procedure which reduces the detrimental effects of the spatial variation of the magnetic environment. 

The approach presented in this work is suitable for real-world applications, where the ability of  scanning the sensor over the target object is typically a requirement. The technique does not require any background subtraction, thus further validating its applicability to to a variety of screening scenarios, from security to industrial monitoring and biomedicine.

We thank Roberto Cecchi (University of Siena) for sharing with us his circuit of the photodiode which inspired our current design.

This work has been partially funded by the Future Aviation Security Solutions (FASS) programme, a joint Department for Transport and Home Office initiative, under contract ACC6008813. With thanks for the technical oversight and programme management provided by Defence Science and Technology Laboratory (Dstl).
This research has also received funding from the UK Engineering and Physical Sciences Research Council (EPSRC) [Grants No. EP/P510270/1 (A.V.), EP/R512400/1 (Y.C.) and EP/N509590/1 (B.M.)].

The data that support the findings of this study are available from the corresponding author upon reasonable request.

\section{\label{sec:ref}References}
\bibliographystyle{aipnum4-1}

\begin{thebibliography}{999}
%
\bibitem[Griffiths(2001)]{griffiths2001}
H. Griffiths,
Magnetic induction tomography.
Meas. Sci. Technol. {\bf 12}, 1126 (2001).
%
 \bibitem[Darrer(2015a)]{brendan2015a}
 B. J. Darrer, J. C. Watson, P. Bartlett, F. Renzoni,
 Magnetic Imaging: a New Tool for UK National Nuclear Security.
Sci. Rep. {\bf 5},  7944 (2015).
%
\bibitem[Alzeibak(1993)]{alzebaik1993}
S. Alzeibak and N.H. Saunders, A feasibility study of in vivo electromagnetic imaging. 
Phys. Med. Biol.  {\bf 38}, 151–160 (1993).
%
\bibitem[Merwa(2004)]{grazfeasibility} 
R. Merwa, K. Hollaus, O. Bir\'o, H. Scharfetter, 
Detection of brain oedema using magnetic induction tomography: a feasibility study of the likely sensitivity and detectability. 
Phys. Meas. {\bf 25}, 347-354 (2004).
%
\bibitem[Zolgharni(2010)]{zol2010} 
M. Zolgharni, H. Griffiths, P. D. Ledger, 
Frequency-difference MIT imaging of cerebral haemorrhage with a hemispherical coil array: numerical modelling. 
Physiol. Meas.  {\bf 31}, S111-S125 (2010).
%
\bibitem[Wang(2017)]{wang2017}
L. Wang and A.M.  Al-Jumaily,
Imaging of Lung Structure Using Holographic Electromagnetic Induction.
IEEE Access {\bf 5}, 20313-20318 (2017).
%
\bibitem[Wickenbrock(2013)]{arne2013}
A. Wickenbrock, F. Tricot, and F. Renzoni,
Magnetic induction measurements using an all-optical $^{87}$Rb atomic magnetometer. 
Appl. Phys. Lett. {\bf 103}, 243503 (2013).
%
\bibitem[Wickenbrock(2014)]{arne2014}
A. Wickenbrock, S. Jurgilas, A. Dow, L. Marmugi, F. Renzoni,
Magnetic induction tomography using an all-optical $^{87}$Rb atomic magnetometer.
Opt. Lett. {\bf 39}, 6367 (2014).
%
\bibitem[Marmugi(2016)]{luca2016}
L. Marmugi, F. Renzoni, Optical Magnetic Induction Tomography of the Heart.
Sci. Rep.  {\bf 6}, 23962 (2016).
%
\bibitem[Deans(2016)]{spie2016} 
C. Deans, L. Marmugi, S. Hussain, F. Renzoni,
Optical atomic magnetometry for magnetic induction tomography of the heart.
Proc. SPIE, {\bf 9900}, 99000F (2016).

%
%

%
%
%
\bibitem[Marmugi(2015)]{marmugi2015spie}
L. Marmugi, S. Hussain, C. Deans, and F. Renzoni, 
Magnetic induction imaging with optical atomic magnetometers: towards applications to screening and surveillance.
Proc. SPIE  {\em 9652}, 965209-965209-11 (2015).
%
\bibitem[Deans(2017)]{cameron2017}
 C. Deans,  L. Marmugi, F. Renzoni, Through-barrier electromagnetic imaging with an atomic magnetometer.
Opt. Express 25, 17911 (2017).
%
\bibitem[Deans(2018b)]{cameron2018b}
C. Deans,  L. Marmugi, and F. Renzoni, 
Active Underwater Detection with an Array of Atomic Magnetometers.
Appl. Opt. {\em  57}, 2346 (2018).
%
\bibitem[Deans(2016)]{cameron2016}
C. Deans, L. Marmugi, S. Hussain, F. Renzoni, 
Electromagnetic Induction Imaging with a Radio-Frequency Atomic Magnetometer.
Appl. Phys. Lett. {\bf 108}, 103503 (2016).
%
\bibitem[Wickenbrock(2016)]{arne2016}
A. Wickenbrock, N. Leefer, J.W. Blanchard, and D. Budker, 
Eddy current imaging with an atomic radio-frequency magnetometer.
Appl. Phys. Lett. {\em 108}, 183507  (2016).
%
\bibitem[Marmugi(2019)]{luca2019}
L. Marmugi,  C. Deans, and F. Renzoni, 
Electromagnetic induction imaging with atomic magnetometers: unlocking the low-conductivity regime.
Appl. Phys. Lett.  {\em 115}, 083503 (2019).
%
\bibitem[Jensen(2019)]{jensen2019}
K. Jensen, M. Zugenmaier, J. Arnbak, H. Stærkind, M. V. Balabas, and E. S. Polzik, 
Detection of low-conductivity objects using eddy current measurements with an optical magnetometer.
Phys. Rev. Research  {\em 1}, 033087 (2019).
%
\bibitem[Deans(2020)]{luca2020}
C. Deans, L. Marmugi, and F. Renzoni,
Sub-Sm$^{-1}$ electromagnetic induction imaging with an unshielded atomic magnetometer.
Appl. Phys. Lett. {\em 116}, 133501 (2020).
%
\bibitem[Bevington(2018)]{patrick2018}
P. Bevington, R. Gartman, W. Chalupczak, C. Deans, L. Marmugi, and F. Renzoni, 
Non-Destructive Structural Imaging of Steelwork with Atomic Magnetometers. 
Appl. Phys. Lett. {\em 113}, 063503 (2018).
%
\bibitem[Bevington(2019a)]{npl2019a}
P. Bevington, R. Gartman, and W. Chalupczak, 
Imaging of material defects with a radio-frequency atomic magnetometer.
Rev. Sc. Instr. {\em 90}, 013103 (2019).
%
\bibitem[Bevington(2019b)]{npl2019b}
P. Bevington, R. Gartman, and W. Chalupczak, 
Enhanced material defect imaging with a radio-frequency atomic magnetometer.
J. Appl. Phys. {\em 125}, 094503 (2019).
%
\bibitem[Budker(2007)]{budker} 
D. Budker, M. Romalis,
Optical magnetometry.
Nat. Phys. {\em 3}, 227--234 (2007).
%
\bibitem[Savukov(2005)]{savukov2005}
I. M. Savukov, S. J. Seltzer, M. V. Romalis, and K. L. Sauer, 
Tunable atomic magnetometer for detection of radio-frequency magnetic fields.
Phys. Rev. Lett. {\bf 95}, 063004, (2005).
%
\bibitem[Schwindt (2004)]{cpt}
Peter D. D. Schwindt, Svenja Knappe, Vishal Shah, Leo Hollberg, and John Kitching, Li-Anne Liew and John Moreland,
Chip-scale atomic magnetometer.
Appl. Phys. Lett. {\em 85}, 6409 (2004);
%
\bibitem{fid}
D. Hunter, S. Piccolomo, J. D. Pritchard, N. L. Brockie, T. E. Dyer, and E. Riis,
Free-Induction-Decay Magnetometer Based on a Microfabricated Cs Vapor Cell.
Phys. Rev. Applied {\em 10}, 014002 (2018).
%
%
 \bibitem[Deans(2018c)]{cameron2018c}
C. Deans, L. Marmugi, and F. Renzoni, 
Sub-picotesla widely-tunable atomic magnetometer operating at room-temperature in unshielded environments.
Rev. Sci. Instr.  {\bf 89}, 083111 (2018).
%
%
%

\end{thebibliography}

\end{document}